\documentstyle[aps,pra,epsfig,amssymb,amsfonts,eqsecnum,floats]{revtex}

\title{Propagation of squeezed radiation through amplifying or absorbing
random media}
\author{M. Patra and C. W. J. Beenakker}
\address{Instituut-Lorentz, Universiteit Leiden, P.O. Box 9506, 2300 RA
Leiden, The Netherlands}

\draft
\newcommand{\raus}{{\text{out}}}
\newcommand{\rein}{{\text{in}}}
\newcommand{\abso}{{\text{abs}}}
\newcommand{\ampl}{{\text{amp}}}
\newcommand{\therm}{{\text{th}}}
\newcommand{\total}{{\text{total}}}
\newcommand{\homo}{{\text{homo}}}
\newcommand{\direkt}{{\text{direct}}}
\newcommand{\probe}{{\text{probe}}}
\newcommand{\minimal}{{\text{min}}}
\newcommand{\eins}{\openone}

\newcommand{\deter}[1]{\text{det}(#1)}

\newcommand{\cotanh}{\text{cotanh}}
\newcommand{\cotan}{\text{cotan}}

\newcommand{\spur}{\text{tr}}
\newcommand{\var}{\text{var}}
\newcommand{\reteil}{\text{Re}}
\newcommand{\que}{\rho}

\begin{document}

\date{October 1999}
\twocolumn[
\widetext
\begin{@twocolumnfalse}

\maketitle

\begin{abstract}
We analyse how nonclassical features of squeezed radiation (in
particular the sub-Poissonian noise) are degraded when it is
transmitted through an amplifying or absorbing medium with randomly
located scattering centra. Both the cases of direct photodetection and
of homodyne detection are considered. Explicit results are obtained
for the dependence of the Fano factor (the ratio of the noise power
and the mean current) on the degree of squeezing of the incident state,
on the length and the mean free path
of the medium, the temperature, and on the 
absorption or amplification rate.
\end{abstract}


\vspace{0.5cm}

\narrowtext

\end{@twocolumnfalse}
]

\section{Introduction}

Squeezed radiation is in a state in which one of the quadratures of
the electric field fluctuates less than the
other~\cite{walls:94,mandel:95}. Such a nonclassical state is useful,
because the fluctuations in the photon flux can be reduced below that
of a Poisson process --- at the expense of enhanced fluctuations in
the phase. Sub-Poissonian noise is a delicate feature of the
radiation, it is easily destroyed by the interaction with an absorbing
or amplifying medium~\cite{henry:96a}. The noise from spontaneous
emission events is responsible for the degradation of the squeezing.

Because of the fundamental and practical importance, there exists a
considerable literature on the propagation of squeezed and other
nonclassical states of light through absorbing or amplifying media. We
cite some of the most recent papers on this
topic~\cite{leonhardt:93a,jeffers:94a,schmidt:96a,barnett:98a,artoni:99a,knoell:99a,alkader:99a}.
The main simplification of these investigations is the restriction to
systems in which the scattering is one-dimensional, such as parallel
dielectric layers. Each propagating mode can then be treated
separately from any other mode. It is the purpose of the present paper
to remove this restriction, by presenting a general theory for
three-dimensional scattering, and to apply it to a medium with
randomly located scattering centra. 

Our work builds on a previous paper~\cite{patra:99a}, in which we
considered the propagation of a coherent state through such a random
medium. Physically, the problem considered here
is different because a coherent state
has Poisson noise, so that the specific nonclassical features of
squeezed radiation do not arise in
Ref.~\onlinecite{patra:99a}. Technically, the difference is that a
squeezed state, as most other nonclassical states, lacks a diagonal
representation in terms of coherent
states~\cite{walls:94,mandel:95}. 
We cannot therefore directly extend the theory of
Ref.~\onlinecite{patra:99a} to the propagation of squeezed states. The
basic idea of our approach remains the same: The photodetection
statistics of the transmitted radiation is related to that of the
incident radiation by means of the scattering matrix of the
medium. The method of random-matrix theory~\cite{beenakker:97a} is
then used to evaluate the noise properties of the transmitted
radiation, averaged over an ensemble of random media with different
positions of the scatterers.

The outline of this paper is as follows. In Sec.~\ref{SecSetup} we
first summarise the scattering formalism, and then show how 
the characteristic function of the
state of the transmitted radiation can be obtained from
that of the incident state. This
allows us to compute the photocount
statistics as measured in direct detection (Sec.~\ref{SecDirektMessung}),
and in homodyne photodetection measurements (Sec.~\ref{SecHomoMessung}).
The expressions in Secs.~\ref{SecSetup}--\ref{SecHomoMessung} are generally
valid for any incident state. 
In Sec.~\ref{SecSqueezed} we specialise to the case that
the incident radiation is in an ideal squeezed
state
(also known as a squeezed state of minimal uncertainty, or as a two-photon
coherent state~\cite{walls:94,mandel:95}). 
The statistics of direct and homodyne
measurements are expressed in terms of the degree of squeezing of the
incident state. The
Fano factor, introduced in Sec.~\ref{SecFano}, quantifies the degree
to which the squeezing has been
destroyed by the propagation through an amplifying or absorbing
medium. The ensemble average of the Fano factor is then computed
using random-matrix theory 
in Sec.~\ref{SecRandom}. We conclude in Sec.~\ref{Zusammenfassung}.

\section{Scattering formulation}
\label{SecSetup}

We consider
an amplifying or absorbing disordered medium embedded in a waveguide
that supports $N(\omega)$ propagating modes at frequency
$\omega$. The conceptual advantage of embedding the medium in a waveguide is
that we can give a scattering formulation in terms of a finite-dimensional
matrix.
The outgoing radiation in mode $n$ is described by an annihilation
operator $a_n^{\raus}(\omega)$, using the convention that modes
$1,2,\ldots,N$ are on the left-hand-side of the medium and modes
$N+1,\ldots,2 N$ are on the right-hand-side. The vector $a^{\raus}$
consists of the operators $a_1^{\raus},a_2^{\raus},\ldots,a_{2
N}^{\raus}$. Similarly, we define a vector $a^{\rein}$ for incoming
radiation.

These two sets of operators each satisfy the bosonic
commutation relations
\begin{equation}
	[a_n(\omega),a_m^\dagger(\omega')]= \delta_{nm}\delta(\omega-\omega')
	,\;
	[a_n(\omega),a_m(\omega')] = 0\;.
	\label{commrel}
\end{equation}	
They are related by the
input-output relations~\cite{jeffers:93a,gruner:96a,beenakker:98a}
\begin{mathletters}
\label{basiceq}
\begin{eqnarray}
        a^{\raus}(\omega) &=& S(\omega) a^{\rein}(\omega) +
               Q(\omega) b(\omega) \;, \\
        a^{\raus}(\omega) &=& S(\omega) a^{\rein}(\omega) +
                V(\omega) c^\dagger(\omega) \;,
\end{eqnarray}
\end{mathletters}
where the first equation is for an absorbing medium and the second for an
amplifying medium.
We have introduced the $2N\times2N$ scattering matrix $S$,
the $2N\times2N$ matrices $Q$ and $V$, and the vectors $b$ and $c$ of
$2 N$ bosonic operators. The scattering matrix
can be decomposed into four $N\times N$ reflection and
transmission matrices,
\begin{equation}
        S = \left( \begin{array}{cc} r' & t' \\ t & r
        \end{array} \right) \;.
\end{equation}
Reciprocity imposes the conditions $t' = t^T$, $r = r^T$,
and $r'={r'}^T$.

The operators $b$ and $c$ account for spontaneous emission in the
medium. They satisfy the bosonic commutation
relations~(\ref{commrel}), hence
\begin{equation}
	Q Q^\dagger = \eins - S S^\dagger,\;
	V V^\dagger = S S^\dagger - \eins\;.
	\label{vvss}
\end{equation}
Their expectation values are
\begin{mathletters}
\label{cexpval}
\begin{eqnarray}
\langle b^\dagger_n(\omega) b_m(\omega')\rangle
        &=& \delta_{nm} \delta(\omega-\omega') f(\omega,T) \;,\\
\langle c_n(\omega) c_m^\dagger(\omega')\rangle
        &=& -\delta_{nm} \delta(\omega-\omega') f(\omega,T) \;.
\end{eqnarray}
\end{mathletters}
The Bose-Einstein function
\begin{equation}
f(\omega,T)=\left[\exp(\hbar\omega/k T)-1\right]^{-1}
\end{equation}
is evaluated at positive temperature $T$ for an absorbing medium
and at negative temperature for an amplifying medium.

It is convenient to discretise the frequency in infinitesimally small steps of
$\Delta$, so that $\omega_p=p\Delta$, and treat the frequency index $p$ as a
separate vector index (in addition to the mode index $n$). For example,
$a^{\raus}_{np}=a^{\raus}_n(\omega_p)$ and
$S_{np,n'p'}=S_{n n'}(\omega_p)\delta_{p p'}$.

The state of the outgoing radiation is described by the characteristic function
\begin{equation}
	\chi_{\raus}(\eta) = \langle : \exp\left[
		\Delta^{1/2} ( a^{\raus\dagger} \eta - \eta^* 
	a^{\raus} ) \right] : \rangle \;,
	\label{xiraus}
\end{equation}
where $\langle : \cdots : \rangle$ indicates the expectation value of a normally
ordered product of operators $a^{\raus}$ and $a^{\raus\dagger}$ (creation
operators to the left of the annihilation operators). The vector $\eta$ has
elements $\eta_{np}$. The density operator of the outgoing radiation is uniquely
defined by the characteristic function $\chi_{\raus}$~\cite{walls:94}.
Similarly, the incoming state has characteristic function
\begin{equation}
	\chi_{\rein}(\eta) = \langle : \exp\left[
		\Delta^{1/2} ( a^{\rein\dagger} \eta - \eta^* a^{\rein} )
	\right]
	: \rangle \;.
	\label{xirein}
\end{equation}
The characteristic function of the thermal radiation inside an absorbing medium
is given by
\begin{eqnarray}
	\chi_{\abso}(\eta) &=& \langle : \exp\left[
		 \Delta^{1/2} (b^\dagger \eta - \eta^* b) \right] :
	\rangle \nonumber\\
	&=& \exp\left( -\sum_{n,p} \eta_{np}^* f(\omega_p,T)
	\eta_{np} \right) \equiv \exp( -\eta^* f \eta ) \;.
	\nonumber\\
	\label{xiabso}
\end{eqnarray}
For an amplifying medium, replacing $b$ by $c^\dagger$ and normal ordering by
anti-normal ordering, one finds instead
\begin{equation}
	\chi_{\ampl}(\eta) = \exp( \eta^* f \eta )\;.
	\label{xiampl}
\end{equation}

Combination of Eqs.\ (\ref{basiceq}) and (\ref{vvss}) with
Eqs.\ (\ref{xiraus}--\ref{xiampl}) yields a relationship between the
characteristic functions of the incoming and outgoing states,
\begin{equation}
	\chi_{\raus}(\eta) = \exp\left( - \eta^*
		(\eins 
		- S S^\dagger) f \eta \right) \chi_{\rein}(S^\dagger \eta) \;.
	\label{chirela1}
\end{equation}
This relation holds both for absorbing and amplifying media, because the
difference in sign in the exponent of Eqs.\ (\ref{xiabso}) and (\ref{xiampl}) is
cancelled by the difference in sign between $Q Q^\dagger = \eins - S S^\dagger$
and $V V^\dagger = - (\eins - S S^\dagger)$.


\begin{figure}[b!]
\centering
\epsfig{file=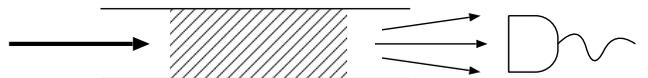,width=3.3in}
\caption{Schematic illustration of direct detection:
Radiation is incident on a random medium (shaded).
The transmitted radiation is absorbed by a photodetector.}
\label{figphoto}
\end{figure}

\section{Photocount distribution}
\label{secPhotocount}
\label{SecDirektMessung}

The photocount distribution is the probability $P(n)$ that $n$ photons are
absorbed by a photodetector within a certain time $\tau$ (see
Fig.~\ref{figphoto}). The factorial
cumulants $\kappa_j$ of $P(n)$ 
(the first two being $\kappa_1=\overline{n}$ and $\kappa_2=
\overline{n(n-1)}-\overline{n}^2$)
are most easily obtained from the generating
function~\cite{mandel:95}
\begin{equation}
F(\xi)=\sum_{j=1}^{\infty} \frac{\kappa_j \xi^j}{j!}
        = \ln\left(\sum_{n=0}^{\infty} (1+\xi)^n P(n) \right)\;.
        \label{Fxieq1}
\end{equation}

The generating
function is determined by a normally ordered expectation
value~\cite{glauber:63a,kelley:64a},
\begin{equation}
	e^{F(\xi)}=\langle:e^{\xi W}:\rangle,\;
	W=\int_0^\tau d t \sum_{n=1}^{2 N} d_n a_n^{\raus\dagger}(t)
	a_n^{\raus}(t) \;.
	\label{Fxieq2}
\end{equation}
Here $d_n\in[0,1]$ is the detection efficiency of the $n$-th mode and the
time-dependent operators are defined as
\begin{equation}
	a_n^{\raus}(t)=(2\pi)^{-1/2} \int_0^{\infty} d\omega\,e^{-i\omega t}
	a_n^{\raus}(\omega)\;.
\end{equation}
Discretising the frequencies as described in Sec.~\ref{SecSetup}, one can write
\begin{equation}
	W = \frac{\Delta^2}{2\pi} \int_0^\tau d t \sum_n d_n \sum_{p,p'}
	e^{i\Delta(p-p')t} a_n^{\raus\dagger}(\omega_p)
	a_n^{\raus}(\omega_{p'})\;.
\end{equation}
This expression can be simplified in the limit $\tau\to\infty$ of long counting
times, when one can set $\Delta=2\pi/\tau$ and use
\begin{equation}
	\int_0^\tau e^{i \Delta (p-p')t} d t = \tau \delta_{p p'}\;.
\end{equation}
Hence, in the long-time limit the generating function is given by
\begin{eqnarray}
	e^{F(\xi)} &=& \bigl\langle : \exp\left( \xi \Delta
	\sum_{n,p} d_n a_n^{\raus\dagger}(\omega_p) a_n^{\raus}(\omega_p) \right)
	: \bigr\rangle \nonumber\\
	&\equiv& 
	\langle : \exp( \xi \Delta a^{\raus\dagger} {\cal D} a^{\raus} ) : \rangle
	\;,
	\label{langfxi1}
\end{eqnarray}
where we have defined the matrix of detector efficiencies
${\cal D}_{np,n'p'}=d_n \delta_{n n'}\delta_{p p'}$. 

Comparing Eqs.\ (\ref{xiraus}) and (\ref{langfxi1}) we see that the generating
function $F(\xi)$ can be obtained from the characteristic function $\chi_{\raus}$
by convolution with a Gaussian,
\begin{equation}
	e^{F(\xi)} = \frac{1}{\deter{-\xi \pi {\cal D}}}
	\int d \eta\,\chi_{\raus}(\eta) \exp\left( \frac{1}{\xi} \eta^* {\cal D}^{-1}
	\eta \right)\;,
	\label{gaussianintro}
\end{equation}
where $\int d\eta$ is an integration over the real and
imaginary parts of $\eta$. We now substitute the relation 
(\ref{chirela1}) between $\chi_{\raus}$ and $\chi_{\rein}$, to
arrive at a relation between $F(\xi)$ and $\chi_{\rein}$:
\begin{eqnarray}
	e^{F(\xi)} &=& \frac{1}{\deter{-\xi\pi{\cal D}}} \int d
	\eta\,\chi_{\rein}(S^\dagger \eta) \nonumber\\
	& & \mbox{~~~}\times \exp\left( \frac{1}{\xi} \eta^*
	{\cal D}^{-1} \eta - \eta^* (\eins - S S^\dagger) f
	\eta \right)\;.
	\label{tempza}
\end{eqnarray}

The fluctuations in the photocount are partly due entirely to thermal
fluctuations, which would exist even without any incident radiation. If we
denote by $F_{\therm}(\xi)$ the generating function of these thermal
fluctuations, then Eq.\ (\ref{tempza}) can be written in the form
\begin{eqnarray}
	F(\xi) &=& F_{\therm}(\xi) \nonumber\\
	&& {}+ \ln\left[ \frac{1}{\deter{\pi M}} \int
	d \eta\,\chi_{\rein}(\eta) \exp\left( - \eta^* M^{-1}
	\eta \right) \right]\;,
	\nonumber\\
	\label{Fxiresulta}\\
	F_{\therm}(\xi) &=& -\ln \det\bigl[ \eins - \xi {\cal D}(\eins - S
	S^\dagger) f \bigr]\;.
	\label{Fthermresult}
\end{eqnarray}
We have defined the Hermitian matrix
\begin{equation}
	M = - \xi S^\dagger [ \eins - \xi {\cal D} ( \eins - S S^\dagger ) f
	]^{-1} {\cal D} S \;,
	\label{mhermitian}
\end{equation}
and we have performed a change of integration variables from $\eta$ to
$S^\dagger \eta$ [with Jacobian $\deter{S S^\dagger}$].

The expression
(\ref{Fthermresult}) generalises the result of Ref.\ \cite{beenakker:98a} to
arbitrary detection-efficiency matrix ${\cal D}$. Returning to a continuous
frequency, it can be written as (recall that $\Delta=2\pi/\tau$)
\begin{eqnarray}
	F_{\therm}(\xi) &=& - \frac{\tau}{2\pi} \int_0^\infty d\omega \ln
	\det\Bigl(\eins \nonumber\\&&{}
	\mbox{~~~~~~~~~~}- \xi D [ \eins - S(\omega) S^\dagger(\omega)]
	f(\omega,T)\Bigr)\;,
\end{eqnarray}
where $D$ is a $2 N \times 2 N$ diagonal matrix containing the
detection efficiencies $d_n$ on the diagonal ($D_{nm}=d_n
\delta_{nm}$). 
The first two factorial cumulants
are
\begin{eqnarray}
	\kappa_1^{\therm} &= &
	\tau \int_0^\infty \frac{d\omega}{2\pi} f(\omega,T) \spur\,D
	[ \eins - S(\omega) S^\dagger(\omega) ] \;, 
	\label{kappa1eq1therm} \\
\kappa_2^{\therm} &=& 
	\tau \int_0^\infty \frac{d\omega}{2\pi}
	f^2(\omega,T) \spur \left(
	D [ \eins - S(\omega) S^\dagger(\omega) ] \right)^2 \;.
\end{eqnarray}	
Note that all factorial cumulants depend linearly on the detection time $\tau$
in the long-time limit.

\begin{figure*}
\centering
\epsfig{file=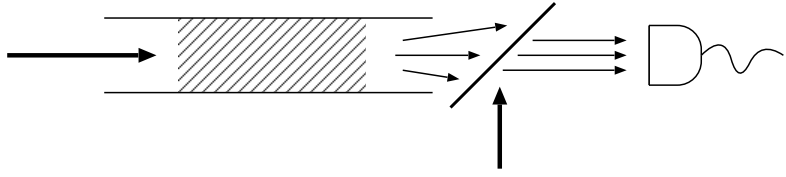}
\caption{Schematic illustration of homodyne detection: At the left,
radiation is incident on a random medium (shaded). At the right, a 
strong coherent beam
is superimposed onto the transmitted radiation, and the combined radiation is
absorbed by a photodetector.}
\label{fighomo}
\end{figure*}

If only the $N$ modes at one side of the waveguide are detected
(with equal efficiency $d$), then $d_n=0$ for $1\le n \le N$ and
$d_n=d$ for $N+1\le n\le 2 N$, hence
\begin{eqnarray}
	F_{\therm}(\xi) &=& -\frac{\tau}{2\pi} \int_0^\infty d \omega \ln
	\det\Bigl(\eins - \xi d [ \eins - r(\omega) r^\dagger(\omega) 
	\nonumber\\&&\mbox{~~~~~~~~~~~~~~~~~~~~~~}{}- t(\omega)
	t^\dagger(\omega)] f(\omega,T)\Bigr)\;,
\end{eqnarray}
in agreement with Ref.\ \cite{beenakker:98b}. 

The difference $F(\xi)-F_{\therm}(\xi)$ contains the noise from the incident
radiation by itself as well as the excess noise due to beating of the incident
radiation with the vacuum fluctuations. If the incident radiation is in a
coherent state, then $\chi_{\rein}(\eta)=\exp(\alpha^* \eta -
\eta^* \alpha)$ for some vector $\alpha$ (called the displacement vector)
with
elements $\alpha_{np}=\alpha_n(\omega_p)$. Substitution into Eq.\
(\ref{Fxiresulta}) gives the generating function
\begin{eqnarray}
	F(\xi)&=&F_{\therm}(\xi) - \alpha^* M \alpha \;,\nonumber\\
	&=& F_{\therm}(\xi) + \frac{\tau \xi}{2\pi} \int_0^\infty
	d \omega\,\alpha^*(\omega) S^\dagger(\omega)\times
	\nonumber\\&&\times
	\Bigl[ \eins - \xi D [ \eins -
	S(\omega) S^\dagger(\omega) ] f(\omega,T) \Bigr]^{-1} D S(\omega)
	\alpha(\omega)
	\nonumber\\
	\label{tempzb}
\end{eqnarray}
The first two factorial cumulants are
\begin{eqnarray}
\kappa_1 &=& 
	\tau \int_0^\infty \frac{d\omega}{2\pi} \alpha^*(\omega)
	S^\dagger(\omega) D S(\omega) \alpha(\omega)
	+ \kappa_1^{\therm} \;,
	\label{kappa1eq1} \\
\kappa_2 &=& 
	2 \tau \int_0^\infty \frac{d\omega}{2\pi} f(\omega,T) \alpha^*(\omega)
	S^\dagger(\omega) D [ \eins - S(\omega) S^\dagger(\omega) ]
	\nonumber\\&&\mbox{~~~~~~~~~~~~~~~~~~~~~~~~~~~~} \times
	D S(\omega) \alpha(\omega) 
	+ \kappa_2^{\therm} \;.
	\label{kappa2eq1}
\end{eqnarray}

If the incident coherent 
radiation is in a single mode $m_0$ and monochromatic with
frequency $\omega_0$, then Eqs.\ (\ref{tempzb}--\ref{kappa2eq1})
simplify for detection in
transmission to
\begin{eqnarray}
F(\xi) &=& F_{\therm}(\xi) \nonumber\\
	\lefteqn{+ \tau \xi d I_0 \Bigl[ t^\dagger \Bigl( \eins
	- \xi d [ \eins - r r^\dagger - t
	t^\dagger ] f(\omega_0,T) \Bigr)^{-1} t \Bigr]_{m_0 m_0}}
	\nonumber\\ \\
\kappa_1 &=& I_0 \tau d [t^\dagger t]_{m_0 m_0} + 
	\kappa_1^{\therm}
	\;, 
	\label{kappa1eq2} \\
\kappa_1^{\therm} &=& \tau d \int_0^\infty \frac{d\omega}{2\pi}
	f(\omega,T) \spur[\eins -r(\omega)
	r^\dagger(\omega) - t(\omega) t^\dagger(\omega)]\;,
	\nonumber\\
	\label{kappa1eq2t}\\
\kappa_2  &=& 2 I_0 \tau d^2 f(\omega_0,T) \left[t^\dagger(\eins-
	r r^\dagger - t t^\dagger)
	t \right]_{m_0 m_0} + \kappa_2^{\therm}\;,
	\nonumber\\\label{kappa2eq2} \\
\kappa_2^{\therm} &=& \tau d^2\int_0^\infty \frac{d\omega}{2\pi} f^2(\omega,T)
	\spur [ \eins - r(\omega) r^\dagger(\omega) 
	\nonumber\\ & & \mbox{~~~~~~~~~~~~~~~~~~~~~~~~~~~~~~~~~~~}{}- t(\omega)
	t^\dagger(\omega) ]^2 \;.
	\label{kappa2eq2t}
\end{eqnarray}
Here $I_0=(2\pi)^{-1} \int_0^\infty d \omega\,|\alpha|^2$
is the incident photon flux and the matrices $r$ and $t$ without
frequency argument are to be evaluated at frequency $\omega_0$.
These are the results of Ref.\
\cite{patra:99a}.


\section{Homodyne detection}
\label{SecHomoMessung}

The photocount measurement described in Sec.~\ref{secPhotocount} 
(known as direct detection)
cannot
distinguish between the two quadratures of the electric field. Such
phase dependent information can be
retrieved by homodyne detection, i.e. by superimposing 
a strong probe beam (described by operators $a^{\probe}$)
onto the signal beam (see Fig.~\ref{fighomo}). The total
radiation incident on the detector is described by the operator
\begin{equation}
	a^{\total} = \kappa^{1/2} a^{\raus} + (1-\kappa)^{1/2} a^{\probe}\;,
\end{equation}
where the factor $\sqrt{\kappa}$ accounts for 
the attenuation of the signal beam by the beam splitter that superimposes it onto
the probe beam. (For simplicity we assume a real scalar $\kappa$, more generally
$\kappa$ would be a complex coupling matrix.)

The characteristic function of $a^{\total}$ is
the product of the characteristic functions of $a^{\raus}$ and $a^{\probe}$.
We assume that the probe beam is in the coherent state 
with displacement vector $\beta$.
From Eq.~(\ref{chirela1}) one gets
\begin{eqnarray}
	\chi_{\total}(\eta) &=& \exp\Bigl[ -\kappa \eta^*
                (\eins - S S^\dagger) f \eta 
                \nonumber\\&&{}
                +(1-\kappa)^{1/2} ( 
                \beta^* \eta -
                \eta^* \beta 
                )\Bigr] \chi_{\rein}(S^\dagger
                \kappa^{1/2} \eta) \;.
\end{eqnarray}                
The generating function $F_{\homo}(\xi)$ of the
photocount distribution in homodyne
detection is given by [cf. Eq.~(\ref{langfxi1})]
\begin{eqnarray}
	\exp[F_{\homo}(\xi)] &=& \langle : \exp( \xi \Delta a^{\total\dagger}
	{\cal D}
	a^{\total} ) : \rangle \\
	&\approx& 
	\exp\bigl[\xi (1-\kappa)\beta^* {\cal D} \beta\bigr]
	\nonumber\\&&{}\times
	\langle : \exp\Bigl( \Delta^{1/2} 
	\sqrt{\kappa(1-\kappa)} \xi 
	\nonumber\\&&\mbox{~~~~~}{}\times
	\bigl[a^{\raus\dagger} {\cal D}
	\beta
	+\beta^* {\cal D} a^{\raus} \bigr]\Bigr) : \rangle \;.
\end{eqnarray}
In the second approximate equality we have linearised the exponent with respect
to $a^{\raus}$, which is justified if the probe beam is much stronger than the
signal beam. The remaining expectation value has the form of a characteristic
function if we take $\xi$ purely imaginary, so that $\xi^*=-\xi$.
The result is
\begin{eqnarray}
	F_{\homo}(\xi) &=& \xi (1-\kappa)\beta^* {\cal D} \beta
	+ \ln \chi_{\raus}(\sqrt{\kappa(1-\kappa)}
	\xi {\cal D}\beta) \nonumber \\
	&=& \xi (1-\kappa)\beta^* {\cal D} \beta
	\nonumber\\
	&& {}+ \kappa(1-\kappa) \xi^2 \beta^* {\cal D}
	(\eins - S S^\dagger) f {\cal D} \beta 
	\nonumber\\ & &{}+ \ln \chi_{\rein}(
	\sqrt{\kappa(1-\kappa)} \xi
	S^\dagger {\cal D} \beta )\;.
	\label{FhomoAllgemein}
\end{eqnarray}
In the second equation we have substituted the relation (\ref{chirela1}) between
$\chi_{\raus}$ and $\chi_{\rein}$.


\section{Squeezed radiation}
\label{SecSqueezed}

We consider the case that the incident radiation is in the ideal squeezed state
$|\epsilon,\alpha\rangle={\cal C} {\cal S}
|0\rangle$~\cite{walls:94,mandel:95}, obtained from the
vacuum state 
$|0\rangle$ by subsequent action of the squeezing operator
\begin{equation}
	{\cal S} = \exp\left[
		\case{1}{2} \Delta\left( a^{\rein} \epsilon^*
	a^{\rein} - a^{\rein\dagger} \epsilon
	a^{\rein\dagger}\right)\right]
\end{equation}
and the displacement operator 
\begin{equation}
	{\cal C} = \exp\left[\Delta^{1/2}\left( a^{\rein\dagger}\alpha - 
	\alpha^* a^{\rein} \right)\right] \;.
\end{equation}
As in the previous sections, we have discretised the frequency,
$\omega_p=p\Delta$, and used the vector of operators
$a^{\rein}_{np}=a^{\rein}_n(\omega_p)$. The complex squeezing parameters
$\epsilon_n(\omega)={\que}_n(\omega) e^{i \phi_n(\omega)}$ are contained in the
diagonal matrix $\epsilon$ with elements
$\epsilon_{np,n'p'}=\epsilon_n(\omega_p)\delta_{n n'}\delta_{p p'}$. Similarly,
the vector $\alpha$ with elements $\alpha_{np}=\alpha_n(\omega_p)$ contains the
displacement parameters. 

The characteristic function of the incident radiation is given by\
\cite{walls:94,yuen:76a}
\begin{eqnarray}
	\chi_{\rein}(\eta)&=&\exp\left[ \alpha^* \eta - \eta^*
	\alpha - \case{1}{4} \eta ( e^{-i\phi} \sinh{2\que})\eta 
	\right.\nonumber\\&& \left.\mbox{~~~}{}
	- \case{1}{4}
	\eta^*(e^{i\phi}\sinh{2\que})\eta^* - 
	\eta^*(\sinh^2 \que)\eta \right]\;.
\end{eqnarray}
According to Eq.\ (\ref{chirela1}), we thus find for the characteristic
function of the outgoing radiation
\begin{eqnarray}
&&\chi_{\raus}(\eta) = \exp\left( \alpha^* S^\dagger \eta
	- \eta^* S \alpha  - \case{1}{4} \eta S^* ( e^{-i\phi}
	\sinh{2\que}) S^\dagger \eta 
	\right.\nonumber\\
&&\left.- \case{1}{4} \eta^* S ( e^{i\phi} \sinh{2 \que}) S^T \eta^*
	- \eta^* [ f - S( f - \sinh^2 \que ) S^\dagger ]
	\eta \right)\;.
	\nonumber\\
	\label{chiraussqueezed}
\end{eqnarray}

The generating function $F(\xi)$ of the photocount distribution is
obtained from 
$\chi_{\rein}$ by convolution with a Gaussian, cf. Eq.\ (\ref{gaussianintro}).
We find
\begin{equation}
	F(\xi) = F_{\therm}(\xi) - \case{1}{2} \ln \det X 
		- \case{1}{2}
		\left(\begin{array}{c}\alpha^* \\ \alpha \end{array}\right)^T
		X^{-1} \left(\begin{array}{c}
		M \alpha\\ M^* \alpha^* \end{array}\right) 
		\label{tempfxi3}
\end{equation}
where the matrix $X$ is defined in terms of the matrix $M$ by
\begin{eqnarray}
	X &=& \eins + \left(\begin{array}{cc}
	M \sinh \que & - M  e^{i\phi} \cosh \que \\
	- M^* e^{-i\phi} \cosh \que & M^* \sinh \que \end{array}\right)
	\nonumber\\ && \mbox{~~~~~~~~~~~~~~~~~~~~~~~~~~~}\cdot
	\left(\begin{array}{cc} \sinh \que & 0 \\ 0 & \sinh \que \end{array}\right)
	\;.
	\label{tempzd}
\end{eqnarray}
If squeezing is absent, ${\que}=0$, hence $X=\eins$ and 
Eq.~(\ref{tempfxi3}) reduces to the
result~(\ref{tempzb}) for coherent radiation. For a squeezed vacuum 
($\alpha=0$) one has
simply $F(\xi)= F_{\therm}(\xi) - \frac{1}{2} \ln \det X$. 

If the radiation is incident only in mode $m_0$, then we may compute the matrix
inverse and the determinant in Eq.~(\ref{tempfxi3}) explicitly. The matrix
$M(\omega)$ defined in Eq.~(\ref{mhermitian}) may be replaced by its $m_0,m_0$
element,
\begin{equation}
	M_{m_0 m_0}(\omega) \equiv m = - \xi \Bigl[
	S^\dagger\bigl(\eins-\xi D[\eins- S
	S^\dagger] f \bigr)^{-1} D S \Bigr]_{m_0 m_0}
\end{equation}
Note that $m$ is real,
since it is the diagonal element of a Hermitian matrix.
The resulting generating function is
\begin{eqnarray}
&&	F(\xi) =
	F_{\therm}(\xi) \nonumber\\
	&&{}- \case{1}{2} \tau \int_0^{\infty}
	\frac{d\omega}{2\pi} \ln( 1 + 2 m \sinh^2 \que - m^2 \sinh^2 \que ) 
	\nonumber\\
	&&{} - \tau
	\int_0^\infty \frac{d\omega}{2\pi} m |\alpha|^2
	\frac{1}{1+2 m \sinh^2
	\que - m^2 \sinh^2 \que} \times
	\nonumber\\
	&& \mbox{~~~~~~}\times \Bigl(
	 1 + m \sinh \que
	[ \sinh \que + \cosh \que \cos(2 \arg \alpha - \phi)  ] \Bigr)
	\nonumber\\
\end{eqnarray}
The first two factorial cumulants, for detection in transmission, are
\begin{eqnarray}
	\kappa_1 &=& \kappa_1^{\therm} + \tau d \int_0^\infty \frac{d\omega}{2\pi}
	( |\alpha|^2 + \sinh^2 \que) [ t^\dagger t]_{m_0 m_0} 
		\label{kappa1transdirekt} \;, \\
	\kappa_2 &=& \kappa_2^{\therm} + 2 \tau d^2 \int_0^\infty
	\frac{d\omega}{2\pi} (|\alpha|^2+\sinh^2 \que) f 
	\nonumber\\ && \mbox{~~~~~~~~~~~~~~~~~~~~~~~~~~~}
	\times [t^\dagger (\eins - r
	r^\dagger - t t^\dagger ) t]_{m_0 m_0} \nonumber\\
	&& {} + \tau d^2 \int_0^\infty \frac{d\omega}{2\pi}
	[t^\dagger t]_{m_0 m_0}^2
	\bigl[ | \alpha \cosh \que - \alpha^* e^{i\phi} \sinh \que|^2 
	\nonumber\\ &&\mbox{~~~~~~~~~} {}
	- |\alpha|^2
	+ \sinh^2 \que ( \cosh^2 \que + \sinh^2 \que ) \bigr]
		\label{kappa2transdirekt}\;,
\end{eqnarray}
where $\kappa_1^{\therm}$ and $\kappa_2^{\therm}$ are given by
Eqs.~(\ref{kappa1eq2t}) and
(\ref{kappa2eq2t}).

The generating function for homodyne detection follows from
Eqs.~(\ref{FhomoAllgemein}) and
(\ref{chiraussqueezed}),
\begin{eqnarray}
&&	F_{\homo}(\xi) = \xi (1-\kappa)\beta^* {\cal D} \beta 
	\nonumber\\&&\mbox{~~~~~~}{}+
	\xi \sqrt{\kappa(1-\kappa)} ( \alpha^* S^\dagger {\cal D} \beta 
	+ \beta^* {\cal D} S \alpha ) \nonumber\\
	&&\mbox{~~~~~~}{}- \case{1}{4}\xi^2 \kappa(1-\kappa)
	[ \beta {\cal D} S^* ( e^{-i\phi} \sinh 2
	\que ) S^\dagger {\cal D} \beta
	\nonumber\\ &&\mbox{~~~~~~~~~~~~~~~~~~~~~~~~~~~}{}
	+ \beta^* {\cal D} S ( e^{i\phi} \sinh 2 \que ) S^T {\cal D} \beta^* ]
	\nonumber\\
	&&\mbox{~~~~~~}{}+\xi^2 \kappa (1-\kappa)
	 \beta^* {\cal D} [ f - S(f - \sinh^2 \que ) S^\dagger ]
	{\cal D} \beta \;.
\end{eqnarray}
All factorial cumulants except for the first two vanish in the
strong-probe approximation.
We may simplify the generating function by assuming that the signal beam is
incident in a single mode $m_0$ and that the probe beam is also in a single mode
$n_0$. For detection in transmission one then has the factorial cumulants
\begin{eqnarray}
	\kappa_1 &=& \tau d \int_0^\infty \frac{d\omega}{2\pi}
	\Bigl( (1-\kappa)|\beta|^2 
	\nonumber\\ && \mbox{~~~~~~~~~~~~~~~}{}+ 2 
	\sqrt{\kappa(1-\kappa)} \reteil[\alpha
	\beta^* t_{n_0 m_0} ] \Bigr) \;, 
	\label{transkappa1} \\
	\kappa_2 &=& -\tau 
	\kappa(1-\kappa) d^2 \int_0^\infty \frac{d\omega}{2\pi}
	\reteil[{\beta^*}^2 e^{i\phi} t_{n_0 m_0}^2   ] \sinh 2\que
	\nonumber \\
	&&{}+2 \tau
	\kappa(1-\kappa) d^2 \int_0^\infty
	\frac{d\omega}{2\pi} |\beta|^2 \Bigl[
	|t_{n_0 m_0}|^2 \sinh^2 \que 
	\nonumber\\&&\mbox{~~~~~~~~~~~~~~~~~~~~~~}{}
	+ f ( \eins - r r^\dagger - t t^\dagger )_{n_0 n_0} \Bigr]
	\label{transkappa2} \;.
\end{eqnarray}

\section{Fano factor}
\label{SecFano}

For the application of these general formulas we focus our attention on the
Fano factor ${\cal F}$, defined as the ratio of the noise power $P=\tau^{-1}
\var\,n$ and the mean current $\bar{I}=\tau^{-1} \overline{n}$:
\begin{equation}
	{\cal F} = P / \bar{I} = 1 + \kappa_2 / \kappa_1 \;.
\end{equation}
(We have assumed the limit $\tau\to\infty$.)
For coherent radiation ${\cal F}=1$, corresponding to Poisson statistics.
Thermal radiation has ${\cal F}>1$ (super-Poissonian). Nonclassical states,
such as squeezed states, can have ${\cal F}<1$.

We assume that the radiation is incident in a single mode $m_0$ and is detected
in transmission (equal efficiency $d$ per transmitted mode). 
We consider a
frequency-resolved measurement, covering a narrow frequency interval around the
central frequency $\omega_0$ of the incident radiation. The thermal
contributions $\kappa_1^{\therm}$ and $\kappa_2^{\therm}$ may then be neglected,
since they are spread out over a wide frequency range. The incident radiation
has Fano factor ${\cal F}_{\rein}$, measured in direct detection with unit
efficiency. For squeezed radiation, one has
\begin{eqnarray}
	{\cal F}_{\rein} &=& 1 +  \frac{1}{|\alpha|^2 + \sinh^2
        \que} \Bigl[
	| \alpha \cosh \que - \alpha^* e^{i\phi} \sinh \que|^2 
	\nonumber\\&&\mbox{~~~~~~~~~~}{}
	- |\alpha|^2
        + \sinh^2 \que ( \cosh^2 \que + \sinh^2 \que )\bigr]
	\;.
	\label{FanoSqueezed}
\end{eqnarray}

We seek the Fano factor of the transmitted radiation, both for direct
detection (${\cal F}_{\direkt}$) and for homodyne detection
(${\cal F}_{\homo}$). Combining Eqs.~(\ref{kappa1transdirekt}) and
(\ref{kappa2transdirekt}), we find for direct detection
\begin{eqnarray}
&&	{\cal F}_{\direkt} = 1 + d ( t^\dagger t )_{m_0 m_0}
	({\cal F}_{\rein}-1) \mbox{~~~~~~~~~~~~~~~~~~~}\nonumber\\
&&	\mbox{~~~~~~~~~}
	{}+ 2 d\,f(\omega_0, T)
	\frac{[t^\dagger(\eins-r r^\dagger-t t^\dagger) t]_{m_0 m_0}}{
	(t^\dagger t)_{m_0 m_0}} \;.
	\label{FanoDirekt}
\end{eqnarray}
The first term is due entirely to the incident
radiation. It is absent for coherent radiation (because then
${\cal F}_{\rein}=0$). The second term is due to the beating of the incident
radiation with the vacuum fluctuations. It is independent of the incident
radiation and was studied in detail in Ref.~\cite{patra:99a}.	
		
The Fano factor in the strong-probe approximation
($|\beta|\to\infty$)
follows from
Eqs.~(\ref{transkappa1}) and (\ref{transkappa2}), with the result
\begin{eqnarray}
	{\cal F}_{\homo} - 1 &=& 2 d \kappa
	|t_{n_0 m_0}|^2 \sinh^2 \que
	\nonumber\\
	&&{}+ 2 d \kappa f(\omega_0,T)
	 ( \eins - r r^\dagger - t t^\dagger )_{n_0 n_0}	
	\nonumber\\
	&&{}- d \kappa \reteil\bigl[
	e^{i(\phi-2 \arg \beta)} t_{n_0 m_0}^2  \bigr] \sinh 2\que
	\;.
	\label{FanoHomo}
\end{eqnarray}
In the strong-probe approximation, it is independent of $\alpha$
and $|\beta|$. Similarly to Eq.~(\ref{FanoDirekt}), the
first term is entirely due to the incident radiation, vanishing for
coherent radiation ($\que=0$), and the second term is due the
beating with vacuum fluctuations. The additional third term describes
the effect of the phase of the probe beam on the measurement.
Typically, in a measurement one would vary the phase of the probe beam
until the Fano factor is minimised, which occurs when $\arg
\beta=\frac{1}{2}\phi + \arg t_{n_0 m_0}$. The resulting Fano factor
${\cal F}_{\homo}^{\minimal}$ is given by
\begin{eqnarray}
	{\cal F}_{\homo}^{\minimal} &=& 1 - 2 d \kappa
	|t_{n_0 m_0}|^2 e^{-\que} \sinh\que
	\nonumber\\ && {}
	+ 2 d \kappa f(\omega_0,T)
	 ( \eins - r r^\dagger - t t^\dagger )_{n_0 n_0}	
	\;.
	\label{FanoHomoMin}
\end{eqnarray}

\section{Ensemble averages}
\label{SecRandom}

The expressions for the Fano factor given in the previous section contain the
reflection and transmission matrices of the waveguide. These are
$N$-dimensional matrices that depend on the positions of the
scatterers inside the waveguide. 
The distribution of these matrices in an ensemble of disordered
waveguides is described by random-matrix theory~\cite{beenakker:97a}. 
Ensemble averages of moments
of $r r^\dagger$ and $t t^\dagger$ for $N\gg1$ have been computed
by Brouwer~\cite{brouwer:98a}, as a function of 
the mean free path $l$ and the amplification
(absorption) length $\xi_a=\sqrt{D \tau_a}$, where $1/\tau_a$ is the
amplification (absorption) rate and $D=c l / 3$ is the diffusion
constant. It is assumed that both $\xi_a$ and $L$ are small compared
to the localisation length $N l$ but large compared to the mean free
path $l$. Obviously, this requires a large number $N$ of propagating
modes. The relative size of $L$ and $\xi_a$ is
arbitrary. 

As sample-to-sample fluctuations are small for $N\gg1$, we
can take in Eq.~(\ref{FanoDirekt}) the averages of numerator and
denominator separately. 
The dependence on the index $m_0$ of the incident mode drops out on
averaging, $\langle \cdots \rangle_{m_0 m_0} = N^{-1}\langle \spur
\cdots \rangle$.
For an absorbing disordered waveguide, we find
\begin{eqnarray}
{\cal F}_{\direkt} &=& 1 + \frac{4 l d}{3 \xi_a\sinh s}
        ( {\cal F}_{\rein} - 1 )
        + \frac{d}{2} f(\omega_0,T) \mbox{~~~~~~~~~~~~~}
        \nonumber\\
        \lefteqn{\times\left[
        3 - \frac{2 s + \cotanh\,s}{\sinh s}
        - \frac{s\,\cotanh\,s - 1}{\sinh^2 s} + \frac{s}{\sinh^3 s}
        \right] \;}
        \nonumber\\
	\label{Fdirekt1}
\end{eqnarray}
We have abbreviated $s=L/\xi_a$.
In the limit of strong absorption, $s\to\infty$,
the Fano factor approaches the
universal limit~\cite{beenakker:99a} ${\cal F}_{\direkt}=1
+\frac{3}{2}d f$. The Fano factor ${\cal F}_{\rein}$ is given by
Eq.~(\ref{FanoSqueezed}) for an incident squeezed state, but
Eq.~(\ref{Fdirekt1}) is more generally valid for any state of the incident
radiation.

The result for an amplifying disordered waveguide
follows by the replacement $\tau_a\to-\tau_a$, hence $\xi_a\to i\xi_a$:
\begin{eqnarray}
{\cal F}_{\direkt} &=& 1 + \frac{4 l d}{3 \xi_a \sin s}
        ( {\cal F}_{\rein} - 1 )
        + \frac{d}{2} f(\omega_0,T) \mbox{~~~~~~~~~~~~~}
        \nonumber\\ \lefteqn{\times\left[
        3 - \frac{2 s - \cotan\,s}{\sin s}
        + \frac{s\,\cotan\,s - 1}{\sin^2 s} - \frac{s}{\sin^3 s}
        \right] \;.}
        \nonumber\\
	\label{Fdirekt2}
\end{eqnarray}
The Fano factor diverges at the laser threshold $s=\pi$. The function
$f(\omega_0,T)$ now has to be evaluated at a negative temperature. For
a complete population inversion of the atomic states $f\to-1$.

\begin{figure*}
\centering
\epsfig{file=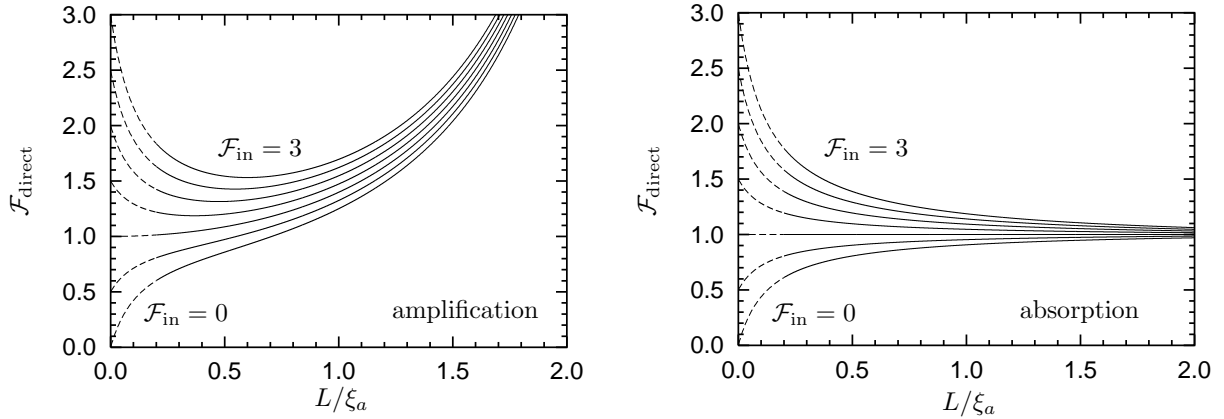,width=6.6in}
\caption{Average Fano factor ${\cal F}_{\direkt}$ for direct detection
as a function of the length of the waveguide. The left panel is for an
amplifying medium [Eq.~(\ref{Fdirekt2}), $f=-1$], the right panel for
an absorbing medium [Eq.~(\ref{Fdirekt1}), $f=10^{-3}$]. In both cases
we took $l/\xi_a=0.1$, $d=1$, and values of
${\cal F}_{\rein}$ increasing from $0$ to
$3$ in steps of $0.5$. The dotted parts of the curves are
extrapolations in the range $L\lesssim l$ that is not covered by
Eqs.~(\ref{Fdirekt1}) and (\ref{Fdirekt2}).
}
\label{fig3}
\end{figure*}

The minimal Fano factor in homodyne detection is given by
Eq.~(\ref{FanoHomoMin}). The average $\langle | t_{n_0 m_0}|^2 \rangle$
is again independent of the mode indices, hence it can be replaced by
$N^{-2} \langle \spur\,t t^\dagger\rangle$. For an absorbing waveguide
we find
\begin{eqnarray}
	{\cal F}_{\homo}^{\minimal} &=& 1
        - \frac{8 l d \kappa}{3 N \xi_a \sinh s} 
	e^{-\que} \sinh\que
	\nonumber\\
        &&\mbox{~~}{}+ \frac{8 l d \kappa}{3 \xi_a} f(\omega_0,T) \left[
        \cotanh\,s + \frac{1}{\sinh s} \right]
	 \;,
	\label{Fhomo1a}
\end{eqnarray}
and for an amplifying waveguide
\begin{eqnarray}
	{\cal F}_{\homo}^{\minimal} &=& 1
        - \frac{8 l d \kappa}{3 N \xi_a \sin s} 
	e^{-\que} \sinh\que
	\nonumber\\
        &&\mbox{~~}{}+ \frac{8 l d \kappa}{3 \xi_a} f(\omega_0,T) \left[
         \cotan\,s - \frac{1}{\sin s} \right] \;.
	\label{Fhomo1b}
\end{eqnarray}
Measurement of the ensemble average ${\cal F}_{\homo}^{\minimal}$ requires
that for every sample the phase of the probe beam is re-adjusted so as to
minimise the Fano factor. If the phase of the probe beam is fixed,
the random phase of $t_{n_0 m_0}$ will average to zero the third term  in
Eq.~(\ref{FanoHomo}). In Eqs.~(\ref{Fhomo1a}) and (\ref{Fhomo1b}) this
amounts to the substitution
$e^{-\que}\to -\sinh\que$.

\begin{figure*}
\centering
\epsfig{file=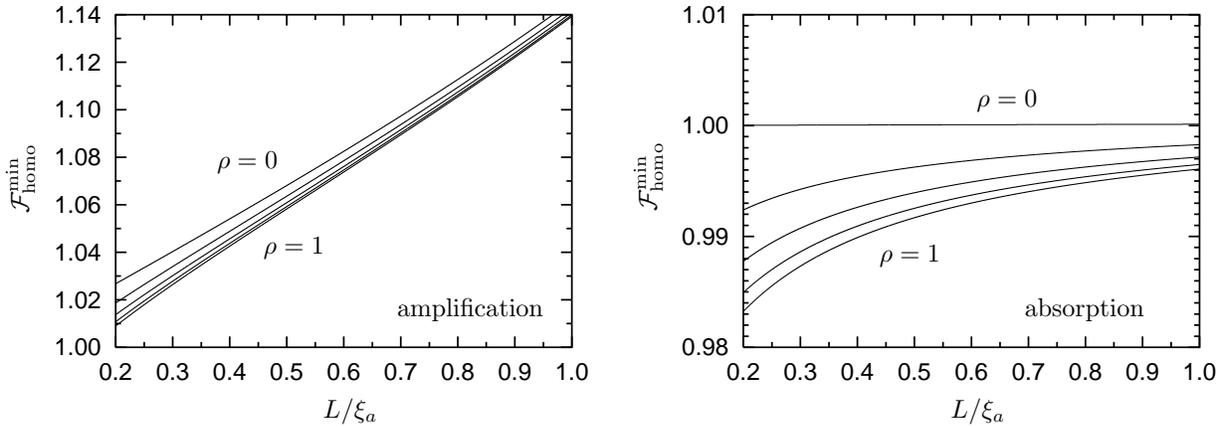,width=6.6in}
\caption{Average minimal Fano factor for homodyne detection,
from Eqs.~(\ref{Fhomo1a}) and (\ref{Fhomo1b}). Same
parameter values as in Fig.~\ref{fig3}, with $N=10$, $\kappa=\frac{1}{2}$,
and $\que$ increasing
from $0$ to $1$ in steps of $0.25$. For $L\lesssim l$ the curves
extrapolate either to $1$ (if $n_0\ne m_0$) or to
$1-e^{-\que}\sinh\que$ (if $n_0=m_0$). (This extrapolation is not shown figure.)
}
\label{fig4}
\end{figure*}

A graphical presentation of the results
(\ref{Fdirekt1})--(\ref{Fhomo1b}) is given in Figs.~\ref{fig3} and
\ref{fig4}. For the absorbing case we have taken $f=10^{-3}$
(corresponding to optical frequencies at $T=3000\,\text{K}$). For the
amplifying case we have taken $f=-1$ (complete population
inversion). The formulas above cannot be used for $L \lesssim l$. The
values of ${\cal F}_{\direkt}$ and ${\cal F}_{\homo}^{\minimal}$ 
for $L=0$ can
be read off from Eqs.~(\ref{FanoDirekt}) and (\ref{FanoHomoMin}),
${\cal F}_{\direkt}\to 1+d({\cal F}_{\rein}-1)$, ${\cal
F}_{\homo}^{\minimal}=1-2 \delta_{n_0 m_0} d \kappa e^{-\que} \sinh\que$.
An extrapolation to $L=0$ is shown dashed in Fig.~\ref{fig3}.

The common feature of the Fano factors plotted in Figs.~\ref{fig3} and
\ref{fig4} is a convergence as the length of the waveguide becomes
longer and longer. For an absorbing medium the $L\to\infty$ limit is
independent of the state of the incident radiation. For an amplifying
medium complete convergence is pre-empted by the laser threshold at
$L=\pi\xi_a$.

\section{Conclusions}
\label{Zusammenfassung}

In conclusion, we have derived general expressions for the
photodetection statistics in terms of the scattering matrix of the
medium through which the radiation has propagated. These expressions
are particularly well suited for evaluation by means of random-matrix
theory, as we have shown by an explicit example, the propagation of
squeezed radiation through an amplifying or absorbing waveguide. The
sub-Poissonian noise that can occur in 
a squeezed state (characterised by a Fano
factor smaller than unity) is destroyed by thermal
fluctuations in an absorbing medium or by spontaneous emission
in an amplifying medium. The theory presented here describes this
interaction of nonclassical radiation with matter in a quantitative
way, without the restriction to one-dimensional scattering of earlier
investigations.

\acknowledgements

This work was supported by the Nederlandse Organisatie voor
Wetenschappelijk Onderzoek (NWO) and the Stichting voor Fundamenteel
Onderzoek der Materie (FOM).


\end{document}